# Multipacting saturation in parallel plate and micro-pulse electron gun

**Authors:** Lang Liao [a) b)], Meng Zhang [a)], Minghua Zhao [a) *]

**Affiliations:** a) Shanghai Institute of Applied Physics, Chinese Academy of Science, Shanghai.
b) University of Chinese Academy of Science, Beijing.
* corresponding author: zhaomh@sinap.ac.cn

**Abstract:** A novel parallel plate model is proposed that divided the electron cloud into three parts at saturation, and it is studied in detail using both an analytical approach and PIC (Particle In Cell) code simulations. As one part of the electron cloud, ribbons modes are suggested by tracking the trajectory of individual particle, and the aim of this mode form is to simplify the progress of multipacting effect in the parallel plate so as to be eliminated by optimizing RF parameters. The micro-pulse electron gun (MPG) has demonstrated the potential to address the need for high peak and average current electron beams, hence studying the multipacting in MPG is essential. On the basis of multipacting studying in the parallel plate, it is clear that increasing the cavity voltage is of interest in yielding high quality beams in the gun.

## 1 Introduction

The multipacting can be initialed with primary electrons at low energy region, and cause an avalanche by largely emitting secondary electrons in the cavity. Hence the multipacting is capable of disturbing the operation of RF structures, such as accelerator cavities and RF power couplers. Since the discovery of the phenomenon of multipacting in 1934 [1], the multipacting has been investigated both theoretically and experimentally in order to suppress it, in most case [2-6]. Under specific circumstances, it was recognized as a new mechanism for amplification of signals [7]. In 1993, the multipacting was used in MPG by taking advantage of its avalanche ability [8].

Previous investigations into multipacting are based on the Cu surface material, from which the average secondary yield $<\delta>$ shake around 1 in the saturation stage [9-10]. The steady sate saturation is characterized by the fact that the number of electrons keep constant with regardless of materials. With the required choice to have a gain of electrons in MPG with a grid-anode, the surface material is important. Previous studies show that MgO material is a good candidate because the maximum secondary emission yield (SEY) of MgO is higher than Cu [11]. Figure 1 shows the secondary emission yield of MgO versus impact energy.

The parallel plate model is much more simpler than any complexes. To get a better feeling for physics before studying factors to affect the mechanism of multipacting, a novel model which divides electron cloud into three parts is proposed in the parallel plate as shown in Fig. 2. It then allowed us to investigate individual behavior of a particle by introducing a PIC code, i.e. VORPAL [12], which allows inclusion of the electromagnetic field and electron distribution minutia. Also a few examples of electron trajectories are presented to demonstrated the generation of hybrid resonance modes occurred in one simulation. Naturally, it is the goal of this part to have a better understanding of the multipacting effect.

Methods used to mitigate the effect of the hybrid modes are also discussed in order to improve the performance of MPG. For single electron or a small group of electron beam, previous studies show that the emission phase would affect the arrival phase [13]. An important practical consideration is the multipacting in MPG will be initialed by dark current, hence it is impossible to identify the initial phase for the initial electrons. In this case one can expects that the cavity voltage is used for optimization only. The cavity voltage is essential to the resonance of multipacting, and increasing cavity voltage leads to increasing the beam quality monitored at the exit of the anode, and the details will be discussed later. Hence the cavity voltage is the one we can handle to boost the energy high enough to balance the effect of space charge force.

## 2 Parallel plate multipacting simulations
### 2.1 Parallel plate model

We consider the motion of electrons in the parallel plate model, from which a sinusoidal voltage of frequency ω apply to the gap with gap distance D. In this paper a sinusoidal-like transverse magnetic field is added in the parallel plate model, hence the situation is similar to the mode in the gun cavity and quite different from the studies before. As a sheet of initial electrons begin to traverse the gap, the competition of RF focusing and charge repulsion becomes more and more fierce. Until the balance of the charge repulsion and the RF focusing, the saturation stage seems to happen with the charge dispersion between the gap.

The parallel plate model we proposed here are benefit from the study of multipacting about Cu material [9-10,13-14]. The difference of multipacting effect between Cu and MgO material mainly reduce to one factor: secondary emission yield curve. The secondary emission yield (SEY) curve, which describe the number of secondary electrons emitted per primary electron, was from Agarwal's work [15]. The SEY model for MgO material with a peak SEY of 20 at 1.5KeV impact energy is shown in Fig. 1. Here we are interested in the saturation stage of MgO material as opposed to the process of saturation, because the saturation stage plays a key role in gun performance.

In this model, the wide bunch which as the result of saturation is artificially divided into three parts. In reality, It is hard to make a distinction between these three parts, but we found that the schedule helps to avoid the influence of collective effects, thereby simplify the analysis process. The electron cloud is assumed to be composed of three parts: a) two-surface impact; b) ribbon modes, as discussed in section 2.3; c) one-surface impact and ping-pong modes. Similar to the parallel plate model, these three parts also exist in the MPG, hence it is essential to discuss the influence in the gun cavity. The part a is contribute to electron gain in the parallel plate as well as the MPG. Especially for MPG, the two-surface multipacting is mandatory to the phase focusing mechanism [4] which is benefit for the generation of micro-pulse. For part c, the one-surface impact and ping-pong modes is commonly existed during the process of multipacting, although it is undesirable in the gun. Ping-pong modes will increase the upper boundary of the multipacting region [14], which makes it hard to be eliminated in the cavity. But there is a possibility to improve the beam quality because of the existence of part b. The part b is necessary to be dressed because of their critical role in affecting the quality of beams and the details about this part will given in section 2.3. The conceive of these three parts will help to give a better understanding of the process of multipacting, thereby help to understand the plasma layer that preventing electrons from being extracted from material by RF field [16], also this model maybe the only way to explain the phenomenon during high power test for MPG that the multipacting effect dysfunction at times [17].

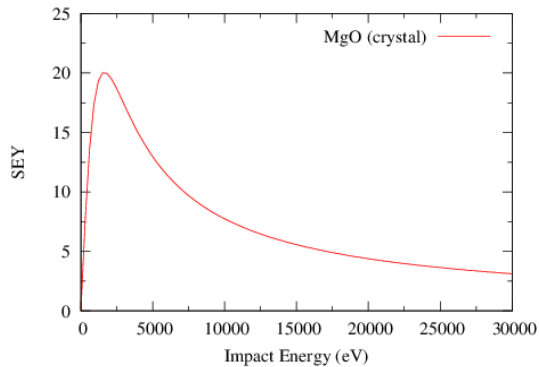

Fig. 1. Secondary emission yield curve for MgO crystal.

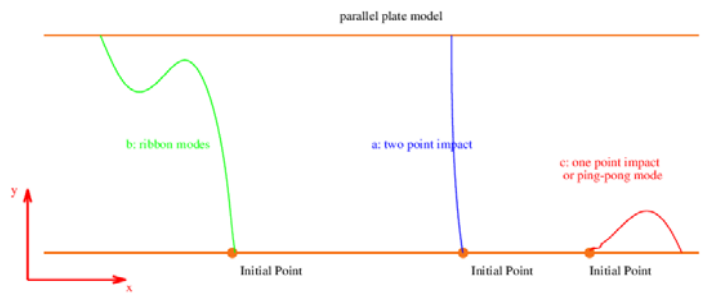

Fig. 2. A schematic drawing of parallel plate multipacting model.

## 2.2  Particle tracking at saturation stage

Instead of investigating collective effect at saturation, it is convenient to track each individual particle in the parallel plate. For the simulations here, after a slice of electrons emitted from one of the electrodes, it is only a few periods for electrons growing to saturation between the parallel plate. Figure 3 shows the phase space at the peak accelerating field of 6.5 MV/m at the given moment of time. An important thing to note from the figure is that the electrons disperse in the parallel plate, which mainly because of space charge effect. As the result of dispersion, the modes in the parallel plate is confused, which, of course, leading to bad understanding.

The trajectories of individual particle are shown in the drawing of Fig. 4. The trajectories in figure express the fact that at saturation lots of multipacting modes are concurrent in the parallel plate. Figure 4 shows the one-surface, 1-st order two-surface and higher order two-surface multipacting respectively. Also there is a new kind of multipacting mode as shown in Fig. 5. A electron emit from one electrode but it never hit the another electrode. It seems like that the electron suspend in the space between upper and lower plate. This kind of mode is contribute to the formation of electron cloud which suspend in the free space. We found that 1-st order two-surface multipacting is commonly existed, which is needed for the running of the MPG, unfortunately the presence of other modes would disturb the operation of electron gun. In what follows, we suggest a resonant form to conclude this "suspension" modes except from 1-st order two-surface multipacting and one-surface multipacting, and try to find a way to kick it out of resonant condition.

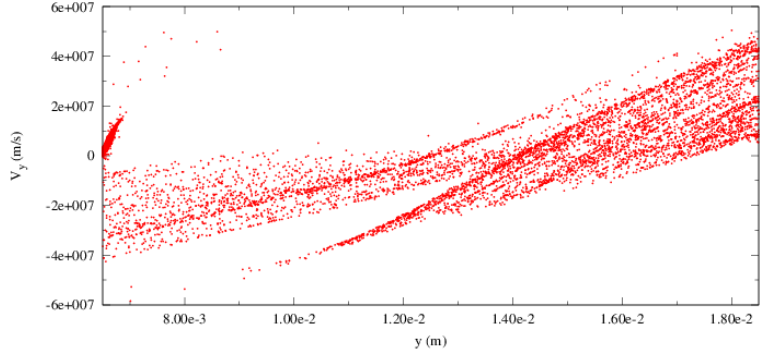

Fig. 3. Velocity in y direction vs y when the peak electric field E0=6.5MV/m in the parallel plate at saturation.

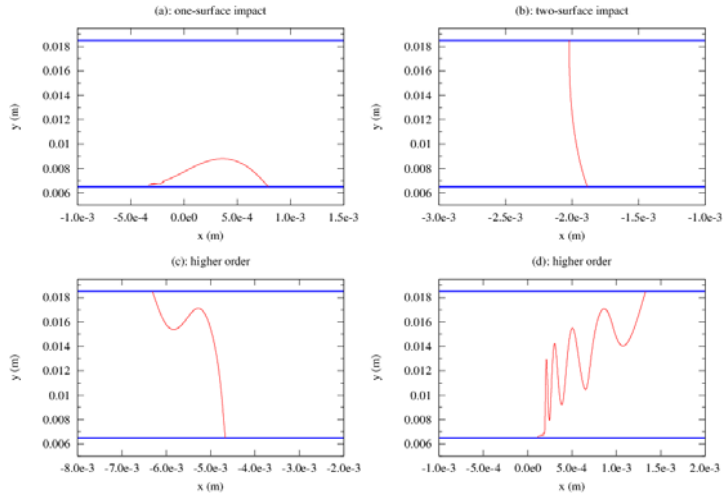

Fig. 4. Electron trajectories for different modes in the parallel plate at saturation, and the gap distance D=12.0mm. (a) One-surface multipacting. (b) 1-st order two-surface multipacting. (c) and (d) Higher order two-surface multipacting which belong to ribbon modes.

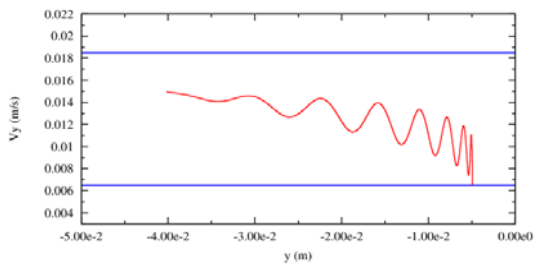

Fig. 5. A kind of electron trajectory in the parallel plate at saturation. The electron emit from one electrode and try to traverse the space. Obviously it will never hit the another electrode, and it seems like that the electron "float" in the free space.

**2.3 Ribbon modes**

Kishek have reported a kind of multipacting mode, namely ping-pong modes [14]. It may seem surprising that the ping-pong modes that combine one-surface and two-surface multipacting within a single period enhance the range of susceptibility of multipacting. In the case, the ping-pong modes extensively exist in the multipacting process. In this part, we will illustrate the

"ribbon modes" that combines higher order modes and suspension-like modes.

In what follows we ignore the space charge effect in our simulations. We assume that electrons emit from one electrode having an initial velocity $v_0$ with an initial phase $\varphi = \theta_0$. The resonant electrons will traverse the gap at the phase of $N\pi + \theta_0$, thereby the resonance equation can be written as

$$y|_{\omega t=N\pi+\theta_0} = D. \quad (1)$$

The situation was addressed in the crossed electric and magnetic fields, so the analytic solution of y in Eq. (1) can be replaced by Eq. (7a) from reference [9]. In this case, the quantity of $\Omega$ is small enough to be ignored, hence the Eq. (1) is easily calculated to be

$$\frac{V_{rf0}}{D} = \frac{m\omega[\omega D^2 - (N\pi+\theta_0)v_0]}{e[\sin(2\theta_0) + \sin\theta_0 + (N\pi+\theta_0)\cos\theta_0]}. \quad (2)$$

Where m/e is the mass-charge ratio, $V_{rf0}$ is the amplitude of RF voltage and $\omega = 2\pi f$ (f is the frequency of cavity). The Eq. (2) can be simplified by integrating the parameters from Table 1, and we have

$$E_0 = \frac{26.4 - 0.086*(\pi+\theta_0)}{\sin(2\theta_0) + \sin\theta_0 + (\pi+\theta_0)\cos\theta_0} \text{ (MV/m)}. \quad (3)$$

Table 1  Parameters used to calculate Eq. (2)

| Parameters | Value |
|---|---|
| Frequency $f$ (GHz) | 2.856 |
| Gap distance D (cm) | 1.2 |
| Initial energy $E_{ini}$ (eV) | 2 |
| Order of multipacting N | 1 |

Where $E_0$ is the axis electric field in the cavity and the relations between $E_0$ and initial phase $\theta_0$ is drawn in Fig. (6). Actually, a fraction of electrons will hit the opposite electrode at unresonance condition. As the electrons hit the electrode before the electric field can be reversed, 1-st order two-surface multipacting is occurred in the model, and this condition corresponds to the trajectories shown in Fig. 7 (a) and (b). But it should be noted that the electrons that traverse the gap after electric field reversed will have different trajectories. In this case, Eq. (3) becomes

$$E_0 < \frac{26.4 - 0.086*(\pi+\theta_0)}{\sin(2\theta_0) + \sin\theta_0 + (\pi+\theta_0)\cos\theta_0} \text{ (MV/m)}. \quad (4)$$

The electrons which meet the requirement of Eq. (4) will yield ribbon-like trajectories as illustrated in Fig. 7 (c), so we termed a series of these modes as "ribbon modes". The A, B and C in Fig. 6 represent electron trajectories corresponding to Fig. 7 (a), (b) and (c) respectively. Actually, the behavior of A, B and C particle are easy to be predicted by Eq. (3) and in turn show a good agreement with Eq. (3).

Let us now consider what is the behavior of ribbon modes. From Fig. 7 (c) it is clear that the ribbon modes not only contains the higher order multipacting electrons, but also the electrons which will never hit the opposite electrode but suspend in the free space of cavity. At the given moment of time when the resonance electrons hit the opposite electrode, unresonance electrons still "float" in the free space, just like a sheet of electron cloud. This kind of suspension behavior will affect the operation of multipacting, thereby affect the quality of electron beams at the gun exit. Fortunately, it is evident that increasing the cavity voltage is a way to eliminate the ribbon modes, hence we will discuss it in MPG.

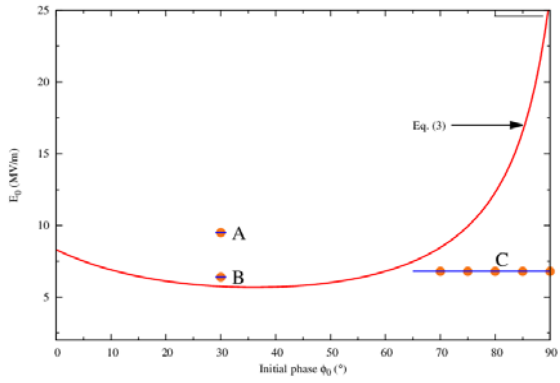

Fig. 6. Drawing of Eq. (2) at the range of [0,90] degrees. Point A represents the electron emit off electrode with initial phase of 30 degree at the electric field of 9.5 MV/m. Point B represents the electron emit with inital phase of 30 degree at the electric field of 6.4 MV/m. Area C represent the electron at the electric field of 6.8 MV/m for different initial phase.

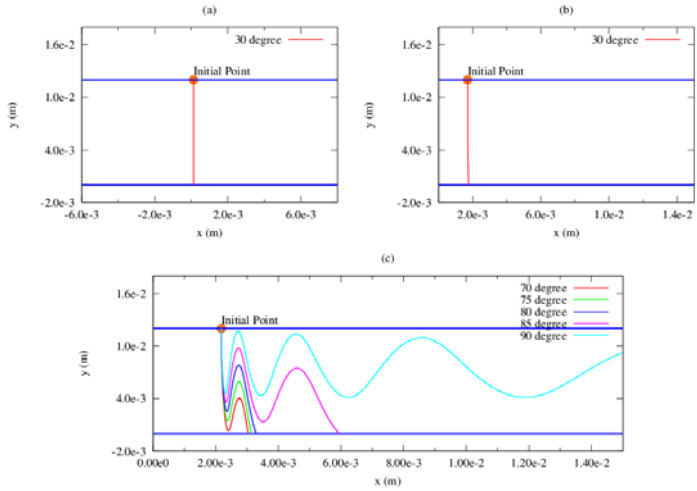

Fig. 7. The trajectories in the parallel plate for different initial phase and electric field. (a) and (b) correspond to 9.5 MV/m and 6.4 MV/m maximum electric field in the model respectively. (c) corresponds to 6.8 MV/m maximum electric field.

**3 Multipacting in the MPG**

From the analysis given previously, it seems that the hybrid modes are concurrent in the process of multipacting. For the requirement of high average-current and high-brightness electron beams of the electron gun, it is benefit to investigate the multipacting effect in the MPG and try to eliminate the effect of ribbon modes, and here we discuss it by constructing a 3-D electron gun model in VORPAL code. Although considering the function of grid and characteristic of cavity shape in the MPG, the results are similar to what we obtained from the parallel plate. A TM010 RF mode motivated in a S-band standing wave cavity is used to accelerate particles. The more details of cavity structure is given in our recent article [18]. The overview of the MPG model is shown in Fig. 8.

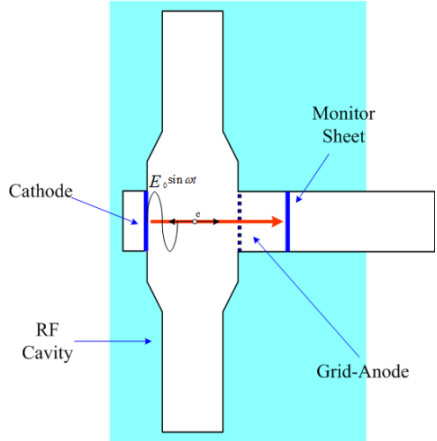

Fig. 8. Schematic of the MPG model.

With the decision to investigate the effect of cavity voltage, we have initiated studies of susceptibility area of multipacting region. The multipacting region, which divided into the first- and higher order and Ping-pong modes, has been drawn as shown in Fig. 9. This kind of susceptibility curve of multipacting is used to figure out the design parameters of MPG, such as the gap distance D and the cavity voltage $V_{rf0}$. From the figure here, the rectangle pink area represents the area of $\delta>2$, where $\delta$ is the secondary emission yield. As the cavity frequency chosen to be a value of 2.856 GHz, the value of D and $V_{rf0}$ can be determined from Fig. 9 by confining modes to the 1-st order two-surface multipacting, but discussion of the details is beyond the scope of this paper here. Hence we will investigate the effect of cavity voltage by setting the gap diatance D to 1.2 cm. The choice of cavity voltage determines the quality of electron beams, and eventually the power level fed into the cavity.

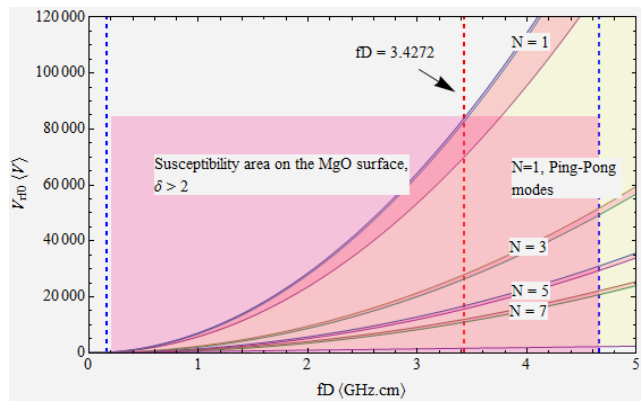

Fig. 9. Susceptibility curve of the hybrid modes. N is the order of two-surface multipacting and normally we chose N=1 for the gun design. The pink area is the susceptibility area on the material surface which corresponds to the property of material.

As a quantifier of beam quality, it is more convenience to evaluate the quality with the bunch length. Figure 10 shows that the micro-pulse structures at saturation for each accelerating field and it is easier to find that the quality of electron beams increases with increasing the cavity voltage. At 5.73 MV/m accelerating field, the resulting beam quality is poor with a bunch length of about 60 percent of RF frequency. The bunch length at 9.5 MV/m accelerating field, which is about 18 ps FHWM, is short enough to be used in the accelerator system. It is important to note that the axis electric field can't increase infinitely, because it should be confined to within the susceptibility area, which depends on the property of surface material. Also from the phase space at saturation as shown in Fig. 11, it is evident that the distribution in Fig. 11 is different from that in Fig. 3. One may conclude that the phase space is changed by increasing the electric field in the cavity so as to eliminate the effect of ribbon modes. Hence increasing the cavity voltage is a appropriate solution to kick ribbon modes out of resonance.

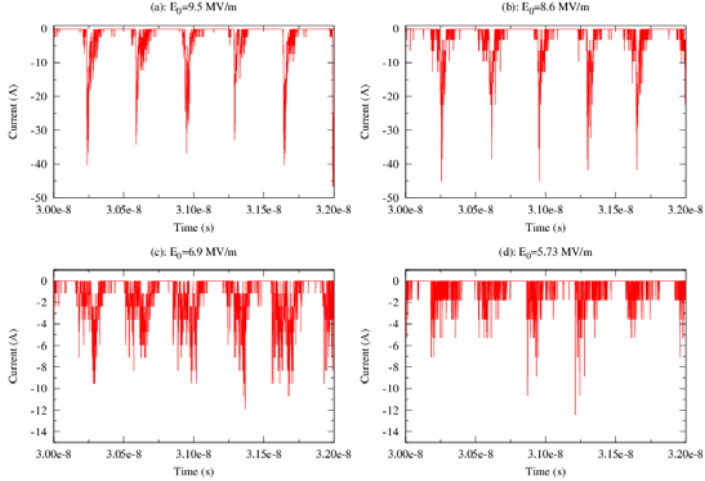

Fig. 10. Micro-bunch structure for different electric fields at saturation stage.

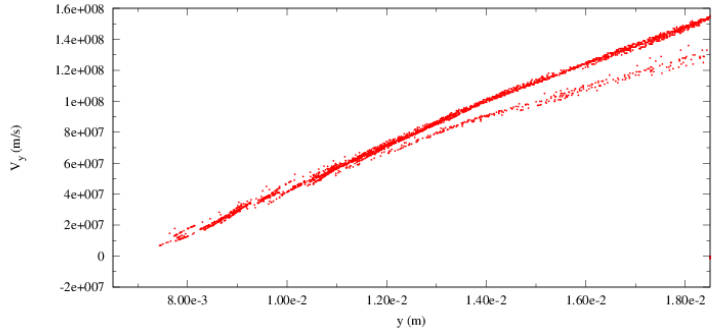

Fig. 11. Velocity in y direction vs y when the peak electric field $E_0$=9.5MV/m at saturation.

## 4 Conclusions

In this paper, multipacting effect is studied both in the parallel plate model and MPG. By constructing a novel model that divided the electron cloud into three parts, it is convenient for us to define a ribbon modes. As a main reason to disturb the operation of MPG, ribbon modes should be focused on in order to eliminate it. Next we carry on simulations in MPG, and it is found that increasing the cavity voltage is benefit to eliminating the effect of ribbon modes, thereby yielding high quality beams at the exit of anode.